# Gyrotropic Bandgaps, Degeneracy Lifting and Selective Suppression of Bloch States in Photonic Systems

Miguel Levy, Amir A. Jalali, Z. Zhou and N. Dissanayake
Physics Department, Michigan Technological University, Houghton, MI 49931

Birefringent magnetophotonic crystals are found to exhibit degeneracy breaking for asymmetric contradirectional coupling in planar waveguides. Fundamental to high-order local normal mode coupling leads to partially overlapping gyrotropic bandgaps inside the Brillouin zone and partial suppression of Bloch mode propagation. A large magneto-optically active reorientation in polarization state is found for allowed Bloch modes at bandgap edges.

**Introduction**

Recent work on one-dimensional Bragg waveguides in gyrotropic systems has revealed the presence of multiple stopbands arising from intermodal coupling and large polarization rotations near the band edges in the gap.[1-3] Polarization rotation enhancement due to photon trapping has also been reported for one-dimensional layered stacks with gyrotropic resonant cavities.[4,5] The latter effect is explained in terms of resonant enhancement in photonic pathlength leading to an increase in net Faraday rotation from multiple passes in a non-reciprocal medium.[4,5] However, the situation in gyrotropic Bragg waveguides, even ones with resonant cavities or half-wave steps, is often fundamentally different because of the presence of linear birefringence. Linear birefringence tends to suppress the Faraday rotation and yet large magneto-optic polarization rotations, responsive to magnetic field reversals, have been reported even for significant non-circular birefringence.[1-3] A different mechanism is at work here. Not photon trapping but the selective suppression of Bloch states in gyrotropic bandgaps. The aim of the present paper is to explain this polarization rotation enhancement mechanism and to present experimental evidence for the existence of gyrotropic bandgap formation and degeneracy lifting inside the Brillouin zone. It is also postulated here that large magneto-optic polarization rotations are possible and realizable even when the Faraday rotation is largely suppressed in the waveguide. Hence, the intriguing possibility exists for magneto-optic polarization rotation devices in waveguide systems even in the presence of significant linear birefringence.

Gyrotropic degenerate bandgaps have recently been predicted for elliptically birefringent magnetophotonic crystals of the type encountered in magnetic garnet waveguide media.[6-8] Elliptical birefringence results from a disparity in refractive indices between elliptically polarized local normal modes, and occurs naturally in planar magnetophotonic crystal waveguides. The possibility to excite different waveguide modes lends added richness to the class of phenomena that can be probed in such systems. Recent work by one of the present authors and co-workers has shown that waveguide modes of different order can be coupled together in one-dimensional magnetophotonic crystals fabricated in bismuth-substituted iron garnet films.[1-3] In particular, bandgaps have been observed where the Bragg reflection mechanism links forward-going fundamental waveguide modes to high-order backscattered ones. The coupling between different waveguide modes travelling in opposite directions leads to particularly interesting phenomena, as discussed in the present paper. Such phenomena include gyrotropic bandgaps where the degeneracy lifting occurs in the wavevector component, the selective suppresion of Bloch states and large changes in Bloch-mode polarization. It should be pointed out that contradirectional coupling of dissimilar modes itself leads to bandgap formation inside the Brillouin zone. The presence of elliptical birefringence, however, leads to degeneracy lifting at the band edges not just in the frequency but also in the wavevector component, and strongly affects the character of the Bloch mode.

**Experimental Background**
The waveguides used in these experiments are single-layer structures formed on bismuth-substituted iron garnet films grown by liquid phase epitaxy on (111) gadolinium gallium garnet (GGG) substrates. These waveguides are multi-moded, the number of modes determined by the film thickness and index contrast between film, cover and substrate. There is a geometrically-induced birefringence, associated with confinement in the waveguide structure, and a stress-induced birefringence as a result of the lattice mismatch between film and substrate. Hence each waveguide mode is associated with a pair of orthogonal polarization states having different effective indices. For the case of gyrotropic waveguides these polarization states are not the standard transverse electric (TE) and transverse magnetic (TM) modes but elliptically-polarized normal modes. The combination of gyrotropy together with geometrical confinement and stress results in a net elliptical birefringence, where the normal modes are elliptically polarized and possess different effective refractive indices.[6,8]

Three sets of samples of different composition were investigated. A 2.86μm-thick $(Bi,Lu)_{2.8}Fe_{4.7}O_{12.1}$ film served as platform for 200μm long gratings with a single phase-shift step in the first set, denoted here as A. Transmittance and polarization rotation measurements on set A were reported by M.Levy and R.Li on a prior publication.[1] Here we analyze these data to extract information about degeneracy lifting, gyrotropic degenerate bandgap formation and the selective suppression of Bloch modes. Waveguide mode refractive indices on this film are 2.318, 2.280, and 2.214 for in-plane polarized light in the slab at a wavelength of 1543 nm. Linear birefringence between in-plane-polarized coupled light and normal-to-the-plane-polarized coupled light were measured at 0.0006, 0.0046, and 0.0106 for the fundamental, first and second-order modes, respectively, with a specific Faraday rotation of 137°/mm.[1] The estimated accuracy in the birefringence data is ±0.0005.

The second set, denoted as set B, consists of 2.7μm-thick $Bi_{0.8}Gd_{0.2}Lu_{2.0}Fe_5O_{12}$ films. Prism coupler refractive index data for TE and TM input polarizations at 1543 nm wavelength on these samples show that the slab waveguides support five modes. Mode indices of the first three, fundamental, first, and second modes were measured at 2.30355, 2.26049, 2.18776, respectively for TE inputs, with linear birefringence of -0.0005 (±0.0005), 0.0036 and 0.0103, respectively. This set had a specific Faraday rotation of 100°/mm. The film composition of the third set, denoted as set C, is $(Bi,Lu,Nd)_3(Fe,Ga,Al)_5O_{12}$, with a thickness of 1.8 µm. Indices on the slab were measured at 2.24601, 2.16210, and 2.02093 for the first three waveguide modes for TE inputs, with birefringence 0.0024, 0.0159, and 0.0312, respectively, all at 1543 nm, and a specific Faraday rotation of 80°/mm.

Ridge waveguides were prepared on these samples by photolithography and plasma etching, with 6µm- and 7µm-wide 800nm-high ridges for sample set A, 7µm-wide ridges and 900nm ridge heights for sample set B and 8µm-wide 600nm-high ridges for sample set C. 200µm-long Bragg reflectors were patterned on the ridges by focused ion beam (FIB) milling as shown in Fig.1. These reflectors had 600nm-deep grooves in set A, 625nm groove depths in sets B and C, and grating periods $\Lambda$ = 338 nm, 335 nm and 346 nm, respectively. A 3.5$\Lambda$ phase shift step was patterned in the middle of the periodic structure for set A, and 10.5$\Lambda$ for set C. No phase shift step was patterned into set B. All samples were treated in a post-patterning cleaning etch of ceric-ammonium nitrate and perchloric acid solution or orthophosphoric acid to remove debris and side-wall damage from the FIB process.

**Scattering Mechanism in the Waveguide**

The tunable laser sources used in these experiments span the range from 1260nm to 1630nm in wavelength, with better than 5pm resolution. Linearly polarized TE inputs were coupled into the waveguide facets. Measurements were taken under a saturation magnetic field of 300 Oe parallel to the waveguide axis in the forward and backward directions. Figures 2 and 3 (right panel) and Fig. 4 plot total transmittance, with contributions from all polarization directions, and polarization rotations obtained from a 360° analyzer scan at each plotted wavelength.

Transmittance measurements on these samples performed by end-fire coupling from a lensed fiber reveal several stopbands. Of particular interest is the Bragg scattering mechanism responsible for the different stopbands. This mechanism has been analyzed and explained before by one of the present authors and co-workers and described in detail in prior publications.[1-3] A key feature of the observed Bragg scattering is the coupling between forward going fundamental waveguide modes and backscattered high-order modes, with different stopbands corresponding to processes with different backscattered modes. The Bragg condition is given by $\lambda = \Lambda(n_f + n_b)$. Here $\lambda$ is the optical wavelength in vacuum, $\Lambda$ the grating period, and $n_f$ and $n_b$ are the modal effective indices of the forward and backward propagating beams, respectively. For surface relief structures such as in Fig. 1 the mode index is an average quantity that depends on film thickness as well as groove depth and shape. The wavelength scan selects different coupling processes, by linking different forward and backwards propagating modes according to the Bragg condition. Each stopband picks out a different scattering process and strongly impacts the character of the Bloch states than can propagate in the grating near the band edges.

The presence of Bragg gratings induces the formation of Bloch states in the periodic structure. These Bloch modes are generally elliptically polarized in gyrotropic materials with non-negligible birefringence, but their polarization state and frequency dispersion characteristics differ from those of the elliptical normal modes that propagate in regions of uniform thickness outside the grating.[1-3, 6-8]

One striking feature of the data is the surprising enhancement in polarization rotation relative to linearly polarized inputs. Polarization rotation is defined here as the angle between the semi-major axis of the output polarization ellipse and the linear input polarization. Typical ellipticities, defined as the ratio of semi-minor to semi-major axes lengths, are plotted in Fig. 5.

Figures 2-4 show that the hybrid coupling of waveguide modes, consisting of a forward-travelling fundamental mode and backscattered high-order mode, yields large polarization rotations that grow with increasing backscattered-mode order. This feature is striking because large rotations are observed even when the presence of non-circular birefringence largely overwhelms the Faraday effect. For uniaxial iron garnets magnetized along the z-direction and negligible absorption, the difference in the diagonal components of the dielectric tensor must be smaller than the gyrotropic component $\varepsilon_{xy}$ for a significant Faraday rotation to manifest itself. The dielectric tensor is given by

$$\vec{\vec{\varepsilon}} = \begin{pmatrix} \varepsilon_{xx} & i\varepsilon_{xy} & 0 \\ -i\varepsilon_{xy} & \varepsilon_{yy} & 0 \\ 0 & 0 & \varepsilon_{zz} \end{pmatrix}, \qquad \text{(Eq. 1)}$$

and $\Delta = (\varepsilon_{yy} - \varepsilon_{xx})/2$ is related to the linear birefringence of the material.

**Polarization Rotations**

The experimental data shown in Figs. 2-3 (right panel) and Fig. 4 contrast with Faraday rotation enhancement due to photon trapping. No significant rotation is seen for the inter-bandgap wavelengths for samples sets A and C, where the light traverses more than 1 mm of material and should exhibit in excess of 80° of Faraday rotation in the absence of linear birefringence. Neither is significant rotation observed at the transmission resonances in the gap for these samples, where resonant pathlength enhancement occurs. In fact, no amount of optical pathlength incrementation through multiple passes will undo the effect of linear birefringence in suppressing the Faraday rotation. However, large magneto-optic rotations, some even greater than 50°, are seen near the stopband edges inside the stopbands in both types of samples. A rather different mechanism from photon trapping is at work to induce these kinds of polarization effects.

In fact, large polarization rotations are also observed in Bragg gratings without resonant cavities, where no photon trapping acts to leverage the optical pathlength. This is seen in Fig. 3 where significant departures from the inter-bandgap rotations are observed near the band edges for set B samples. The measured rotation is not significantly affected by the location of the 200μm-long Bragg reflector on the ridge. Tests were carried out with the gratings both near the center and near the facets along the waveguide ridge axis.

That these rotations are magneto-optic in nature is seen from the response of the samples to magnetization reversals. Figures 2-4 plot the orientation of the semimajor axis of the output ellipse for opposite magnetization directions evincing a clear reversal in the sign of the rotation relative to the input linearly polarized light (right panel). Red and blue data points display polarization rotations for opposite magnetizations in Figs. 2 and 3, as do solid circles and triangles in Fig. 4.

**Gyrotropic Bandgaps and Degeneracy Lifting**

Along with the experimental data for transmittance and polarization rotations, Fig. 2 and Fig. 3 [left panels] plot the computed bands and Bloch-mode semi-major axis orientations for sample sets A and B. A particularly noteworthy feature of these plots concerns the character of the bandgaps for high-order mode backscattering. Bandgap doublets are observed inside the Brillouin zone, slightly displaced from each other both in frequency and wavevector. These are gyrotropic bandgap doublets.[7,8] The fundamental-mode backscattering Bragg condition $\lambda = 2\Lambda n_f$, on the other hand, occurs at the Brillouin zone edge.

A striking feature of these plots is the formation of frequency-shifted gyrotropic bandgap doublets, leading to a selective suppresion of Bloch states at frequencies that lie within one of the gaps in the doublet but outside the other. It is precisely at these wavelengths that a significant rotation in the polarization of the **allowed** Bloch mode is observed. These large rotations are in fact experimentally observed, as can be seen from the experimental data plotted in these figures.

**Theoretical Model**

The model we construct is based on a stack structure with bilayer unit cell, depicted in Fig. 6, that captures many of the essential features of the waveguides under consideration. In particular an alternating system of elliptical birefringent states is introduced in adjacent layers. The Bloch states for this system can be expressed as a linear combination of local normal modes.[9] These states satisfy the Floquet–Bloch theorem through the eigenvalue equation $\vec{\vec{T}}\,\mathbf{E} = \exp(iK\Lambda)\,\mathbf{E}$, where the transfer matrix $\vec{\vec{T}}$ translates the Bloch state by one unit cell. $K$ is the Bloch wave vector, and $\Lambda$ is the period of the periodic structure.[6,10] The use of local normal modes in coupled mode theory is discussed by Dietrich Marcuse in his book on the theory of dielectric optical waveguides.[9] We refer the interested reader to that work for an extensive discussion of this type of mode coupling treatment.

The main elements of the model we present here have been described in detail in prior publications by Levy and Jalali, with one key difference.[6,8] Whereas before forward and backscattered local normal modes did not differ except for propagation direction, here we allow the backward-propagating local normal modes to differ in refractive index and polarization state from the forward-travelling modes. The rationale behind this choice of local normal mode basis rests on the scattering mechanism responsible for the formation of multiple stopbands in one-dimensional magnetophotonic crystal waveguides, a mechanism that connects different waveguide modes through contradirectional coupling.[3]

Thus the Bloch mode in layer $n$ is expressed as follows:

$$\vec{E}(z,t) = \left[\left(E_{01}\exp(i\frac{\omega}{c}n_+^f(z-z_n))\right)\hat{e}_+^f + \left(E_{02}\exp(-i\frac{\omega}{c}n_+^b(z-z_n))\right)\hat{e}_+^b\right]e^{-i\omega t}$$
$$+\left[\left(E_{03}\exp(i\frac{\omega}{c}n_-^f(z-z_n))\right)\hat{e}_-^f + \left(E_{04}\exp(-i\frac{\omega}{c}n_-^b(z-z_n))\right)\hat{e}_-^b\right]e^{-i\omega t}$$

(Eq. 2)

The superscripts *f* and *b* stand for forward and backward, with propagation along the z-direction. In this model, Bloch states are described by a linear combination of underlying local normal modes in each birefringent layer, weighted by amplitudes $E_{0i}$, i =1-4.[6,8] The mode indices $n_\pm^{f,b}$ are assumed to correspond to local waveguide modes of opposite helicity, with birefringence parameters $\Delta = (\varepsilon_{yy} - \varepsilon_{xx})/2$. Thus $n_\pm^2 = \bar{\varepsilon} \pm \sqrt{\Delta^2 + \varepsilon_{xy}^2}$ with $\bar{\varepsilon} = (\varepsilon_{yy} + \varepsilon_{xx})/2$ The gyrotropy parameter $\varepsilon_{xy}$ is equal to the experimentally measured value for the particular film.

A transfer matrix method is used to compute the band structure and polarization rotation. The local dielectric components $\varepsilon_{xx}$ and $\varepsilon_{yy}$ are allowed to deviate from the experimentally measured values for the slab waveguide in order to account for local thickness variations, lateral waveguide confinement and uncertainties in the local normal mode indices. In particular, it is assumed that thickness variations in the grating impact the birefringence and mode indices locally thus inducing different normal mode polarization states in adjacent layers in the model system. Forward and backward modes have different propagation constants and their polarization states are expressed as

$$\hat{e}_\pm^{f,b} = \frac{1}{\sqrt{2}} \begin{pmatrix} \cos\alpha^{f,b} \pm \sin\alpha^{f,b} \\ \pm i\cos\alpha^{f,b} - i\sin\alpha^{f,b} \\ 0 \end{pmatrix}, \quad \text{(Eq. 3)}$$

with $\tan(2\alpha^{f,b}) = \Delta^{f,b}/\varepsilon_{xy}$.[6,8] Deviations up to 2% in modal refractive indices from the measured slab waveguide data, with 0.1% maximum grid size are allowed in order to match stopband bandwidths, center wavelengths and polarization rotations. It should be noted that a difference in polarization states $\hat{e}_\pm$ between adjacent layers is found to be necessary for lifting the degeneracy between Bloch modes.[6,8] This condition was predicted by some of the authors in a previous publication and is found to apply here to unequal forward and backward modes as well.[6,8]

Figures 2 and 3 (left panel) plot the calculated band structure for a stack model system with different local normal modes travelling in opposite directions. Two nearly overlapping gyrotropic degenerate bandgaps or doublet, corresponding to different Bloch states are visible for high-order mode backscattering. The figures also displays the output polarization in the forward direction (calculated and measured), parameterized by the angular orientation of the semi-major axis of the polarization ellipse relative to the input (linear) polarization. The selective suppression of one or another of the Bloch states by the overlapping bandgaps leads to largely enhanced polarization rotations. The experimental transmittance is plotted as well. The presence of a phase shift step in sample set A is accounted for by the model through its effect on the transmittance bandwidth.[11]

Although the polarization rotation outside the stopbands is almost fully suppressed by the large birefringence in sample set A, a non-negligible rotation does occur in sample set B due to the small birefringence of the forward going fundamental mode. The polarization rotation of a waveguide of equal length but without Bragg reflector is taken as a baseline in this case.

The calculated polarization rotations evince an asymmetric double-lobe character around the baseline as exhibited in the experiments. These lobes stem from the polarization response of different Bloch modes, allowed by the gyrotropic bandgap doublets to selectively propagate at different wavelength ranges. The model is able to explain the presence of large magneto-optically switchable rotations near the band edges, their spectral character (the frequencies at which they occur) and the bandwidth and center-wavelengths of the stopbands quite accurately. In addition it elucidates the origin of the large changes in polarization orientation inside the stopbands, showing that these occur at frequencies that lie within the gap in one or another of the Bloch bands but not both. At the same time, the strength of the calculated polarization rotations is occasionally found to be somewhat lower than the experimentally measured values. This is a deficiency in the model that may require a full waveguide-mode treatment or a multilayer (as opposed to bilayer) unit-cell-stack model to correct.

**Conclusions**

The present work examines degeneracy breaking and bandgap formation in gyrotropic systems exhibiting elliptical birefringence, namely, birefringence between elliptically polarized local normal modes. It is found that the coupling between counter-propagating elliptically birefringent local normal modes of different order leads to the lifting of Bloch state degeneracy inside the Brillouin zone. The gyrotropy and birefringence parameters $\varepsilon_{xy}$ and $\Delta$, contribute to this process through the normal mode refractive indices $n_{\pm}$, where $n_{\pm}^2 = \bar{\varepsilon} \pm \sqrt{\Delta^2 + \varepsilon_{xy}^2}$. This process leads to the formation of partially overlapping bandgaps and the selective suppression of Bloch mode propagation inside the gap. A magnetically tunable and large reorientation in the allowed Bloch mode polarization is observed experimentally and calculated theoretically.

**Acknowledgments**
Support by the NSF through grants ECCS-0520814 and DMR-0709669 is acknowledged. ML and AAJ thank A. Merzlikin for illuminating discussions and Horst Dötsch and V.J. Fratello for the virgin iron garnet films.

FIGURE CAPTIONS

Fig. 1. (a) SEM micrograph of one-dimensional waveguide Bragg-filter with phase shift step. (b) SEM micrograph of one-dimensional waveguide Bragg-filter without phase shift step.

Fig. 2 The figure plots the measured transmittance and polarization response [right panel], and calculated band structure and polarization rotation [left panel] of a one-dimensional photonic crystal with phase shift step patterned on a 2.86μm-thick $(Bi,Lu)_{2.8}Fe_{4.7}O_{12.1}$ film (sample set A). Separate curves for the calculated semi-major axis orientation correspond to different Bloch states. The red and blue data points on the right panel describe the orientation of the semi-major axis of the polarization ellipse for opposite magnetizations collinear with the ridge waveguide axis.

Fig. 3 Measured transmittance and polarization response [right panel], and calculated band structure and polarization rotation [left panel] of a one-dimensional Bragg filter without phase shift step patterned on a 2.7μm-thick $Bi_{0.8}Gd_{0.2}Lu_{2.0}Fe_5O_{12}$ film (sample set B). Separate curves for the calculated semi-major axis orientation correspond to different Bloch states. The red and blue data points on the right describe the orientation of the semi-major axis of the output polarization ellipse for opposite magnetizations collinear with the ridge waveguide axis.

Fig. 4 Measured transmittance and polarization response for a one-dimensional photonic crystal with a single phase shift step patterned on a $(Bi,Lu,Nd)_3(Fe,Ga,Al)_5O_{12}$ film of thickness 1.8 μm. Black triangles and solid circles plot the orientation of the semi-major axis of the output polarization ellipse relative to the linear input polarization for opposite magnetzation directions colinear with the ridge waveguide axis.

Fig. 5 Measured ellipticity in the output polarization for Bragg filters patterned on $(Bi,Lu)_{2.8}Fe_{4.7}O_{12.1}$ film (sample set A), $Bi_{0.8}Gd_{0.2}Lu_{2.0}Fe_5O_{12}$ film (sample set B) and $(Bi,Lu,Nd)_3(Fe,Ga,Al)_5O_{12}$ film (samples set C). Ellipticity is defined as the ratio of the semi-minor to semi-major axes of the polarization ellipse in the optical electric field amplitude. The horizontal double-tipped arrows indicate the locations of the stopbands in each case.

Fig. 6 Model structure consisting of a bilayer unit cell of length $\Lambda$. The transfer matrix operator is denoted by **T**. The forward and backward normal mode refractive indices $n_\pm^{f,b}$ are allowed to differ from each other. Similarly, the normal mode indices in adjacent layers (n) and (n+1) are different from each other.

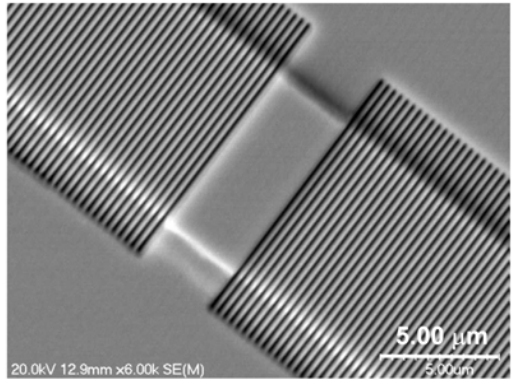

Fig. 1(a)

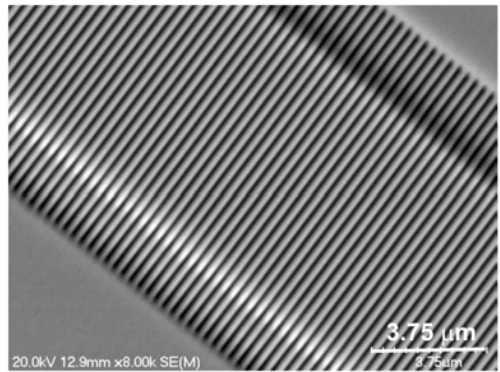

Fig. 1(b)

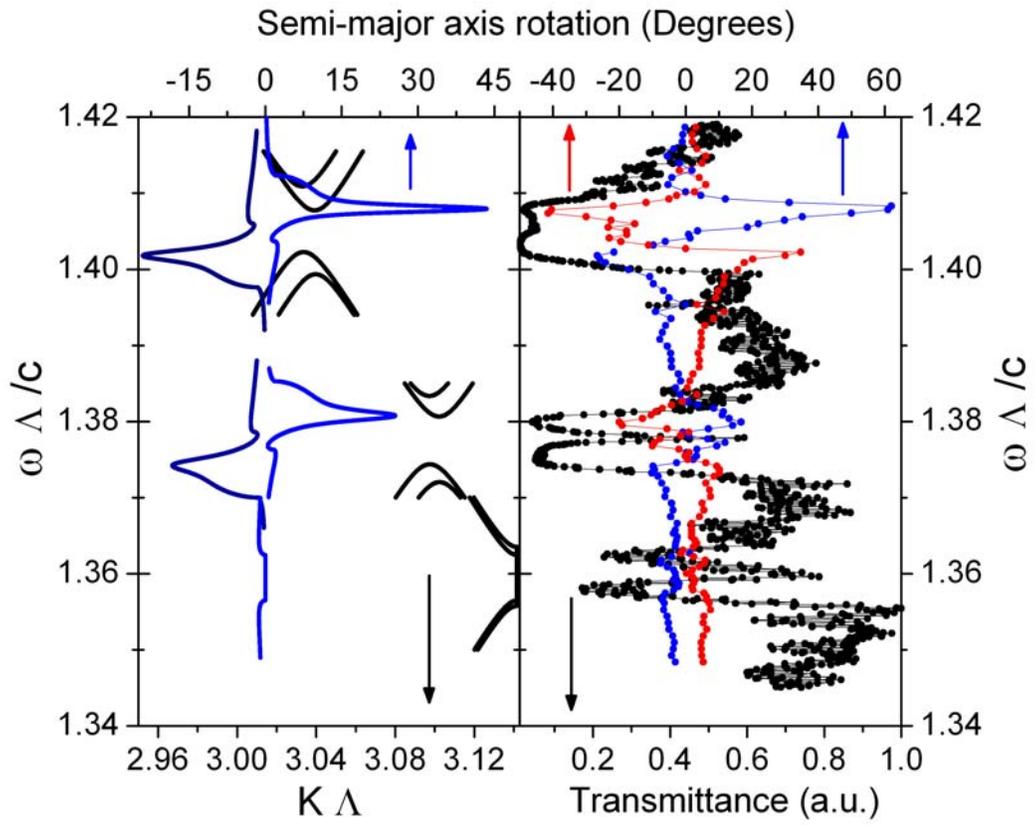

Fig. 2

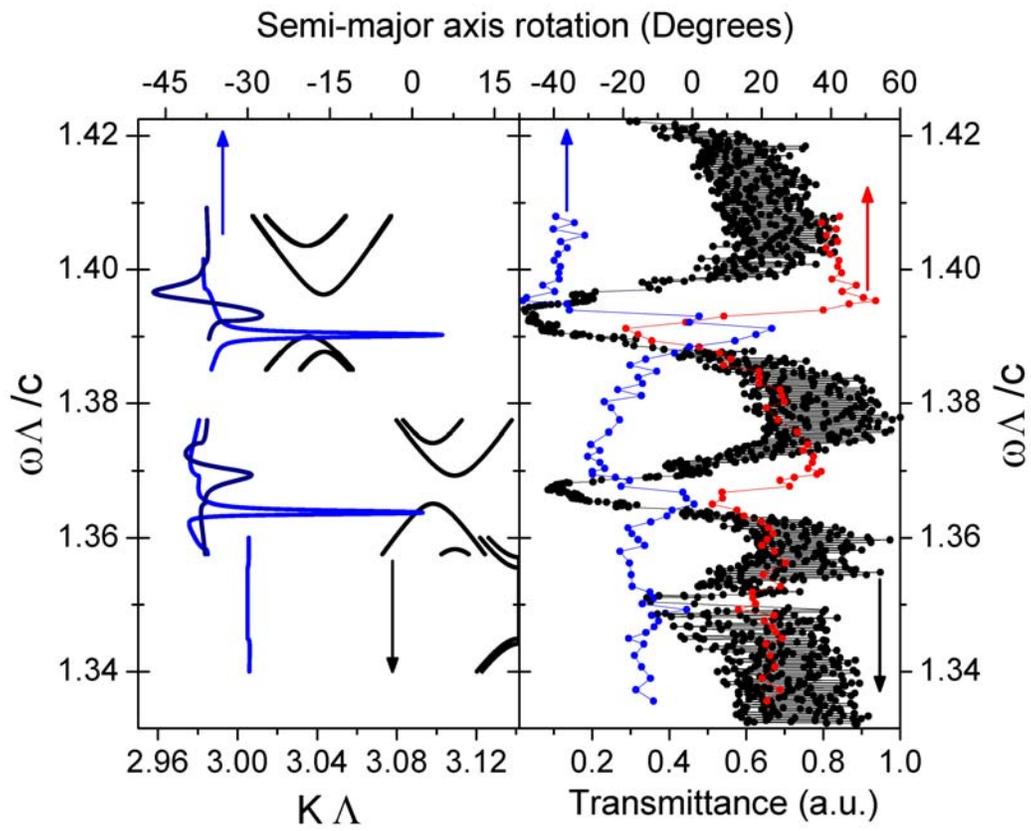

Fig. 3

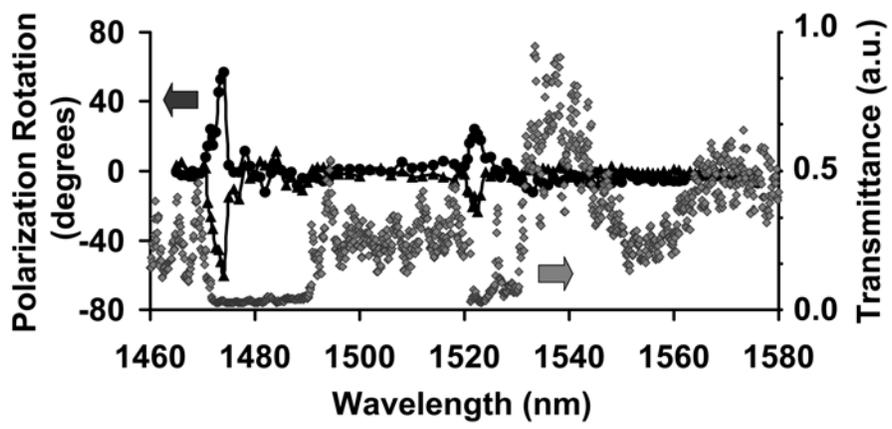

Fig. 4

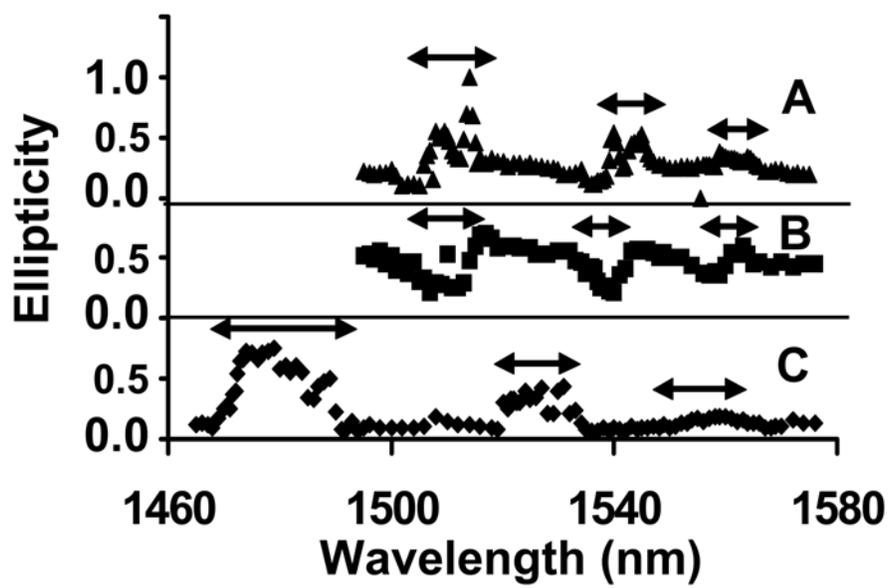

Fig. 5

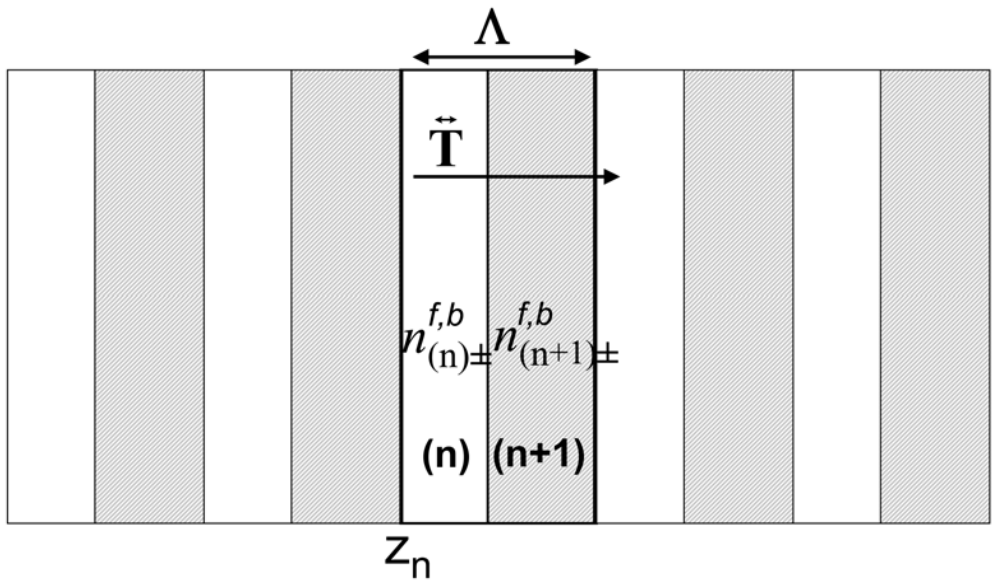

Fig. 6